\documentclass[12pt]{article}

\newcommand \be  {\begin{equation}}
\newcommand \ee  {\end{equation}}

\usepackage{graphicx}

\begin{document}   

\title{Comment on {\em Numerical Study on Aging Dynamics in the $3D$
Ising Spin Glass Model}} 

\author{Enzo Marinari$^{(a)}$, Giorgio Parisi$^{(b)}$ 
and Juan J. Ruiz-Lorenzo$^{(c)}$\\[0.3em]
$^{(a)}$  
{\small Dipartimento di Fisica and INFN, Universit\`a di Cagliari}\\
{\small Cittadella Universitaria, S. P. Monserrato-Sestu Km. 0.7}\\
{\small 09042 Monserrato (CA), Italy}\\
{\small \tt marinari@ca.infn.it}\\[0.3em]
$^{(b)}$  
{\small Dipartimento di Fisica and INFN, Universit\`a di Roma}
{\small {\em La Sapienza} }\\
{\small P. A. Moro 2, 00185 Roma, Italy}\\
{\small \tt giorgio.parisi@roma1.infn.it}\\[0.3em]
$^{(c)}$  
{\small Departamento de F\'{\i}sica Te\'orica I, 
Universidad Complutense de Madrid}\\
{\small Ciudad Universitaria, 28040 Madrid, Spain}\\
{\small \tt ruiz@lattice.fis.ucm.es}\\[0.5em]
}

\date{April 1999}

\maketitle
 
\begin{abstract}
We show that the dynamical behavior of the $3D$ Ising spin glass with
Gaussian couplings is not compatible with a droplet dynamics. We show
that this is implied from the data of reference \cite{KOYOTA}, that,
when analyzed in an accurate way, give multiple evidences of this
fact.  Our study is based on the analysis of the overlap-overlap
correlation function, at different values of the separation $r$ and of
the time $t$.
\end{abstract}


In a very interesting paper \cite{KOYOTA} Komori, Yoshino and Takayama
discuss the dynamical behavior of the three dimensional ($3D$) Edwards
Anderson (EA) Ising Spin Glass with Gaussian couplings. A number of
very accurate numerical evidences are used to try to determine if the
$3D$ EA model behaves in analogy to the Replica Symmetry Broken (RSB)
solution of the Mean Field Sherrington Kirkpatrick model
\cite{PARISI}, or if its behavior is the one typical of coarsening of
domain walls \cite{BRAY}, in agreement with a picture of separated
droplets \cite{DROPLET}.

The evidences presented in \cite{KOYOTA} can seem at first view to
have mixed signs.  For sake of completeness let us try to summarize
all the available evidence.

As it is well explained in the paper \cite{KOYOTA} some of the
findings are obviously devastating for a droplet picture. The free
energy barriers $B_L$ found on a lattice of linear size $L$ do not
grow with $L$ like $L^\psi$, but with a logarithmic dependence
$B_L\propto \log(L)$ (this was already observed by Kisker, Rieger,
Santen and Schreckenberg in \cite{KSSR}). Secondly, at variance with
the droplet picture, the width of the distribution of the free energy
barriers instead of increasing with $L$ turns out to be remarkably
constant.  These two features do not have any problem with the RSB
picture.

Two more findings are not in contradiction with any of the two
exemplar behaviors. The relaxation process has an Arrhenius-type form,
with the characteristic time of the form $\tau=\tau_0
\exp\left(\frac{B}{T}\right)$, and the decay of the energy density to
its infinite time asymptotic value follows a power law. Both the
droplet picture and the RSB scheme imply a power law decay of the
energy density.

The crucial point, as rightly pointed out in \cite{KOYOTA}, is the
dynamical behavior of the overlap-overlap ($q-q$) correlation
functions $G(r,t)$ (computed starting from two different random spin
configurations, considering overlaps at distance $r$ and equal time
$t$). A simple scaling of $G$ as a function of $\frac{r}{\xi(t)}$
would indeed be a strong argument in favor of a droplet behavior,
while evidence for a different scaling points to a RSB pattern. The
analysis of \cite{KOYOTA} claims to show a droplet like behavior of
$G(r,t)$. Here we show that, on the contrary, the numerical data
exclude this simple scaling form.  We show that they behave in the
non-standard way already discussed in reference \cite{MPRR}. We
perform a careful data analysis, and show in detail where the analysis
of \cite{KOYOTA} fails. In this way we show that the observed picture
is fully compatible with an RSB behavior, and that none of the
available numerical supports a droplet-like behavior: one more crucial
piece of evidence, the one concerning the behavior of the
overlap-overlap correlation functions, conjures with the observed
behavior of the free energy barriers to exclude a dynamical droplet
like behavior of $3D$ Ising spin glasses.

We will try to make clear to the reader how delicate this analysis is,
how even a careful treatment like the one of reference \cite{KOYOTA}
can misdirected by only apparent straight lines, and why we can get a
completely safe evidence of a non-trivial scaling of $G(r,t)$. 

As we have already said the crucial difference in the approach to
equilibrium among the droplet approach and the RSB theory is in the
behavior of the $q-q$ correlation function during the approach to
equilibrium.

Let us define

\begin{equation}
  G(r) \equiv \lim_{t\to\infty} G(r,t)\ .
\end{equation}
In the droplet model we have

\begin{equation}
  \lim_{r\to\infty}G(r) = \overline{q} \ne 0\ ,
\end{equation}
while in the RSB solution of the Mean Field theory (in the $q\simeq 0$
sector of the theory: we are analyzing this case since we start out
from equilibrium on a very large lattice) we find~\footnote{A few
details about our runs: we have analyzed four samples of a $3D$ Ising
spin glass with Gaussian couplings, periodic boundary conditions, on a
lattice of linear size $L=64$. We have computed the overlap-overlap
correlation function for runs of time length $t_n=100\cdot 2^n$, where
$n$ runs from $1$ to $13$.  We average the correlation function over
the whole run: this gives a measured $G$ that is numerically slightly
different from the one of \cite{KOYOTA} but has the same scaling
behavior.  We have used random initial configurations.}

\begin{equation}
  G(r) \approx r^{-\mu}\ ,
\end{equation}
and therefore

\begin{equation}
  \lim_{r\to\infty}G(r) = 0\ .
\end{equation}
In reference \cite{MPRR} we have studied this problem and we have
concluded that $\mu \approx .5$, and therefore that a droplet-like
behavior is inconsistent with the results of numerical simulations.

\begin{figure}
\centering
\includegraphics[width=10cm, angle=270]{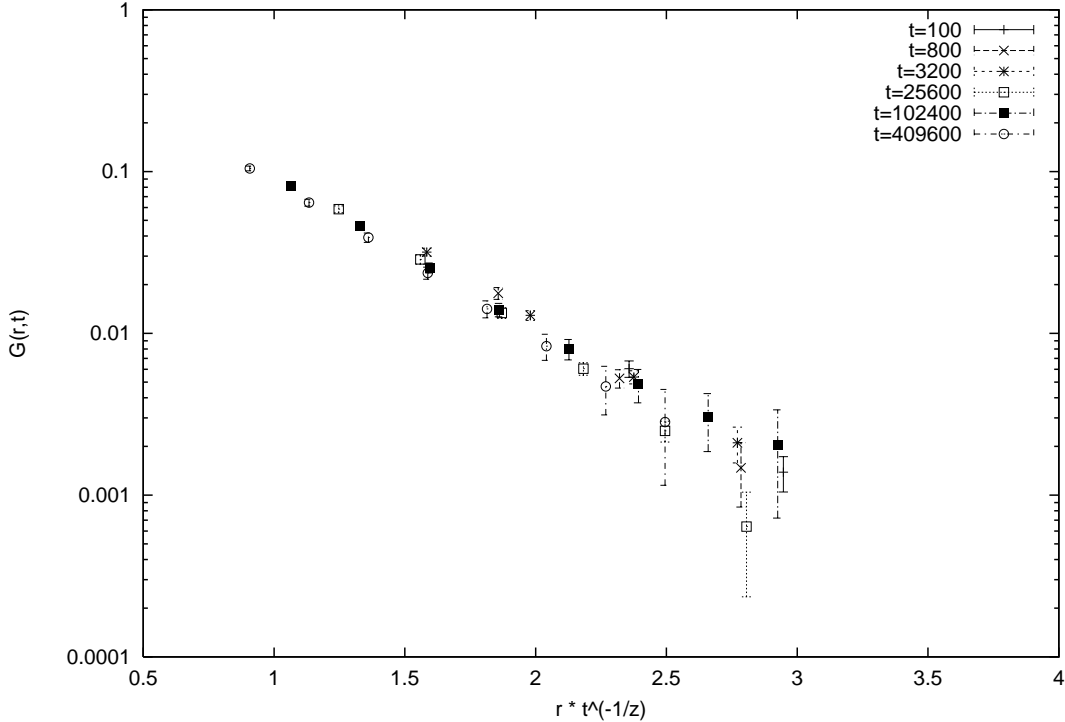}
\caption[1]{The overlap-overlap correlation functions
$G(r,t)$ versus $\frac{r}{\xi}$ for different $r$ and $t$ values.}
\label{F-ALL}
\end{figure}

\begin{figure}
\centering
\includegraphics[width=10cm, angle=270]{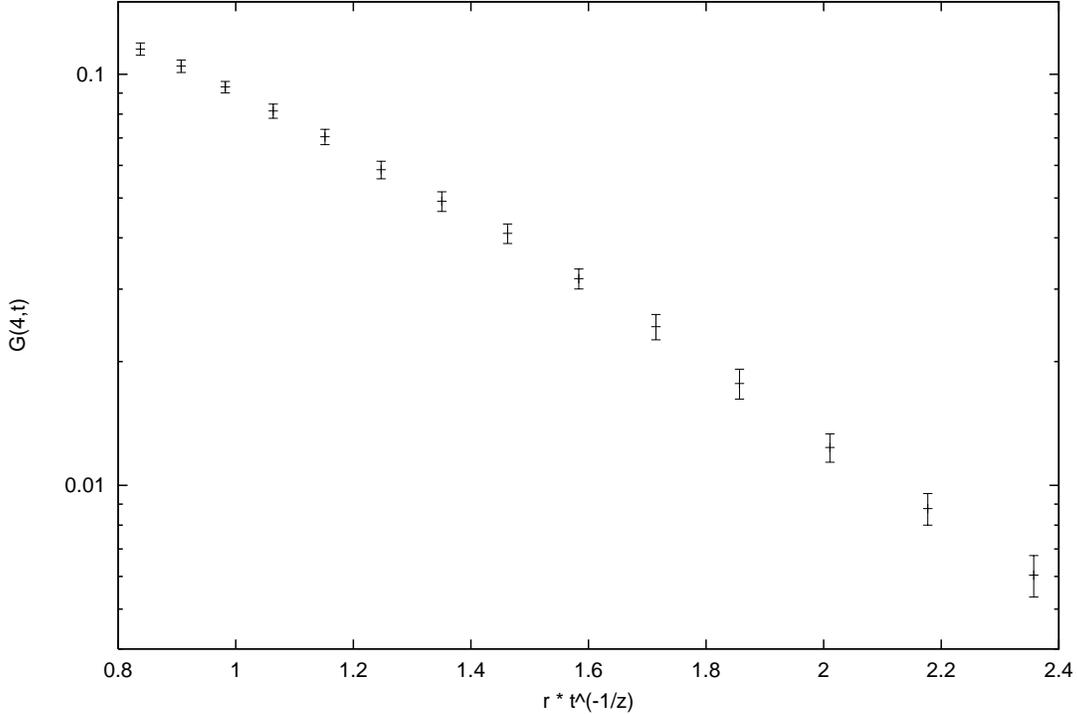}
\caption[1]{$G(r=4,t)$ versus $\frac{4}{\xi}$. These data are not on a
straight line.}
\label{F-R4-KO}
\end{figure}

\begin{figure}
\centering
\includegraphics[width=10cm, angle=270]{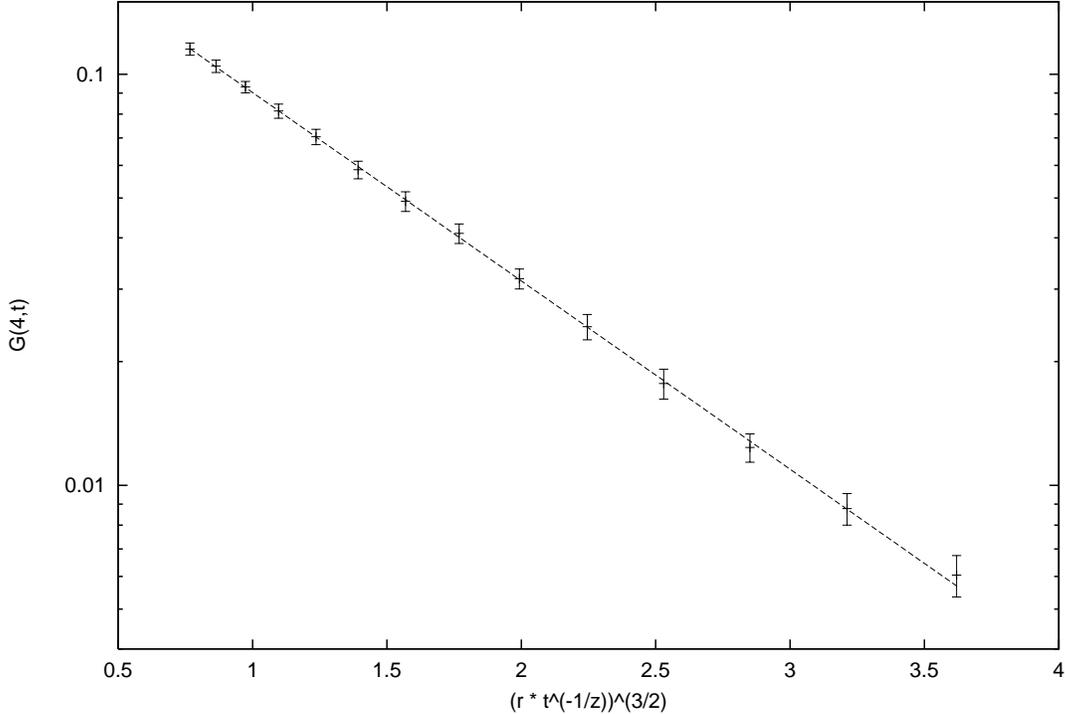}
\caption[1]{$G(r=4,t)$ versus
$\left(\frac{4}{\xi}\right)^{\frac32}$. The dashed curve is a straight line.}
\label{F-R4-OK}
\end{figure}

Reference \cite{KOYOTA} argues that the numerical data for $r\ge 4$,
$T=0.7$, scale like

\begin{equation}
  G(r,t)=M\left(\frac{r}{\xi(t)}\right)\ ,
  \label{PAPER}
\end{equation}
where $\xi(t)=t^{\frac{1}{z(T)}}$, $z(T=0.7)\simeq 8.71$ and $M$ is a
scaling function.  As we have already argued the validity of this
analysis would be a good point in favor of a droplet like behavior of
$3D$ Ising spin glasses (at variance with the many other evidences of
\cite{KOYOTA}, that we have discussed before, that point against the
validity of a droplet picture). Since this conclusion is different
from the one we had reached in \cite{MPRR} we will analyze here its
derivation in better detail. In the rest of the paper (in particular
in figures \ref{F-ALL}, \ref{F-R4-KO} and \ref{F-R4-OK}) we use the
value $z(T=0.7)=8.71$.

The conclusion of \cite{KOYOTA} is based on the plot of the raw data
for $G(r,t)$ as a function of $\frac{r}{\xi(t)}$ for $r\ge 4$ (figure
$5$ of \cite{KOYOTA}). So at first, mainly to exclude the possibility
of programming errors, we compare our raw data (\cite{MPRR,MPR}) to
the ones of figure $5$ of \cite{KOYOTA}. We plot in figure \ref{F-ALL}
our raw data for $G(r,t)$ versus $\frac{r}{\xi}$: by comparing with
\cite{KOYOTA} one can see they are not substantially different, so we
can feel safe about the quality of the numerical data of both
\cite{KOYOTA} and \cite{MPRR,MPR}.  In figure \ref{F-ALL} we have
selected the same $x$ and $y$ scales than in figure $5$ of
\cite{KOYOTA}, to make the comparison of the two sets of data
easier. In order to make a visual comparison easier we have also
selected time values as close as possible to the ones of
\cite{KOYOTA}. In figure \ref{F-ALL} of this note like in figure $5$
of \cite{KOYOTA} the data can seem at first view well aligned on a
straight line, i.e. decreasing with a simple exponential behavior.  We
will show now that this is just not true, and that these data clearly
follow a non-exponential behavior.

The fact that the decay is not purely exponential can be clearly seen,
for example, by plotting only the points with $r=4$ on the same
semilogarithmic scale: we do that in figure \ref{F-R4-KO}. Here it is
clear that the points are not on a straight line. It is just the
dispersion of the many points in figure \ref{F-ALL} that make them
looking like being on a straight line: they are not, and figure
\ref{F-R4-KO} shows it clearly.

We continue our analysis by trying to double check the conclusion of
\cite{MPRR} leading to the functional dependence

\begin{equation}
  G(r,t) = 
  \frac{g}{r^{\frac12}}\ 
\exp\left[-\left(\frac{r}{\xi(t)}\right)^{\frac32}\right]\ .
\end{equation}
We start by plotting in figure \ref{F-R4-OK} the same data of figure
\ref{F-R4-KO} versus $\left( \frac{x}{\xi(t)} \right)^{\frac32}$,
together with the (very good) best fit to the form

\begin{equation}
  G(r) \ \exp{\left(-\frac{A(r)}{\xi(t)^{\frac32}}\right)}\ .
\end{equation}
(at fixed $r$ as a function of $t$ for the two parameters $G$ and
$A$). It is clear from figure \ref{F-R4-OK} and from the quality of
the best fit that this is the scale where the data follow a straight
line, not the one of figure \ref{F-R4-KO}.

\begin{figure}
\centering
\includegraphics[width=10cm, angle=270]{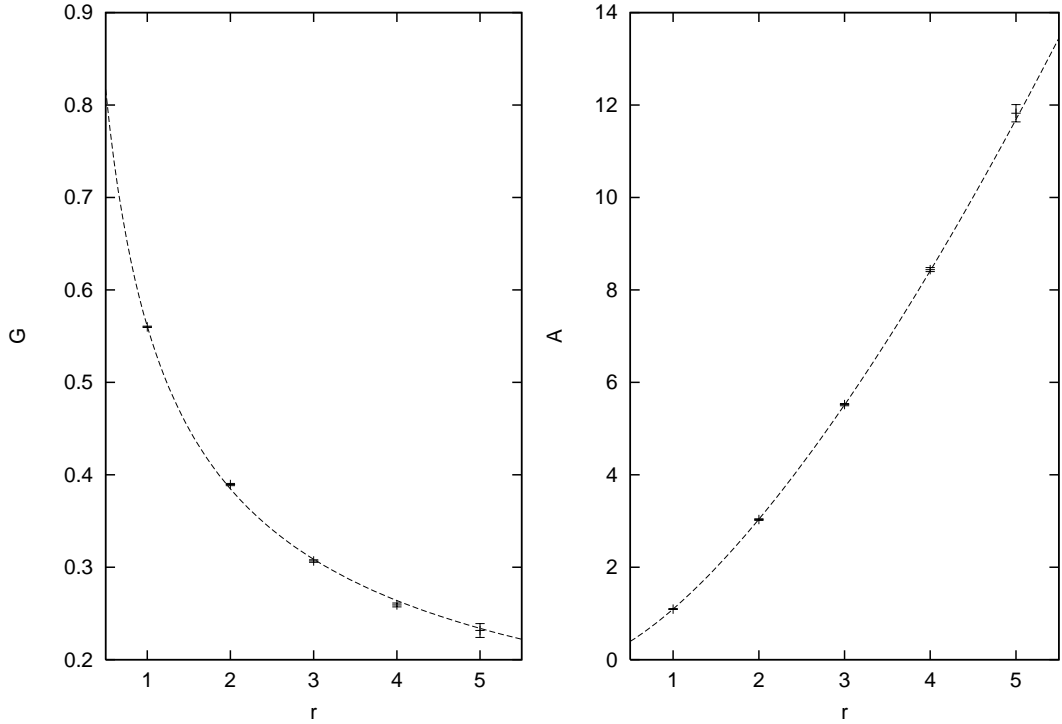}
\caption[1]{$G(r)$ and $A(r)$ versus $r$ and the best fit (dashed
lines) to a power law (see the text for more details). }
\label{F-GA}
\end{figure}

We repeat the procedure of figure \ref{F-R4-OK} for different
distances (finding always a very good best fit), and we fit the
results to a power law (see figure \ref{F-GA}). We fit

\begin{equation}
  G(r) = r^{\epsilon_1}\ , \ \ 
  A(r) = r^{\epsilon_2}\ ,
\end{equation}
and for the best fit $\epsilon_1=-0.54$ and $\epsilon_2=1.47$, in very
good agreement with our initial Ansatz.

\begin{figure}
\centering
\includegraphics[width=10cm, angle=270]{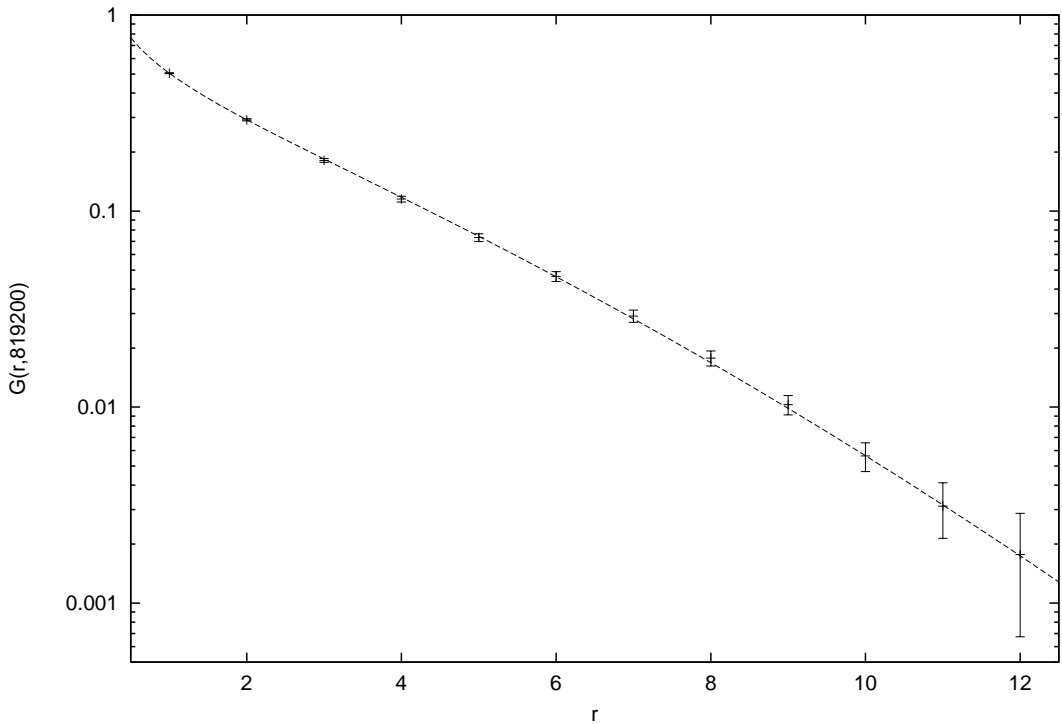}
\caption[1]{$G(r,t=819200)$ versus $r$. The dashed line is our best  fit 
to $\frac{a}{r^{\frac12}}\exp{\left(-b\cdot r^{\frac32}\right)}$. }
\label{F-GT}
\end{figure}

Finally, in figure \ref{F-GT}, we show $G$ as a function of $r$ for
our larger available time ($t=819200$), and the best fit to the form
$a\ r^{-\frac12}\exp{\left(-b\ r^{\frac32}\right)}$.  The
best fit is again very good.  It is interesting to note that, as a
final confirmation of our point, the behavior of the function plotted
in figure \ref{F-GT} is clearly not the one of a simple exponential,
but that in a limited $r$ region it can be mistakenly fitted with a
simple exponential.

We believe that these data and the data of \cite{KOYOTA}, when
correctly analyzed, give strong evidence in favor of a Replica
Symmetry Breaking like dynamical behavior of $3D$ spin glasses, and
that they show beyond any doubt that the droplet model cannot not
describe the numerical data one obtains for the $q-q$ correlation
function $G(r,t)$.

\end{document}